\documentclass[aps,prl,showpacs,floatfix,onecolumn,superscriptaddress,longbibliography]{revtex4-1}

\usepackage[pdftex]{graphicx}
\usepackage{color}
\usepackage{hyperref}
\usepackage{verbatim}
\usepackage{soul}
\usepackage{glossaries}
\usepackage{upgreek}
\usepackage{amsmath}
\usepackage[normalem]{ulem}

\def\TMGO{TmMgGaO$_4$}

\def\Q{\textbf{Q}}
\def\TN{$T_\mathrm{N}$}

\newacronym{CEF}{CEF}{crystal electric field}
\newacronym{SOC}{SOC}{spin-orbit coupling}
\newacronym{FWHM}{FWHM}{full-width at half-maximum}
\newacronym{LSW}{LSW}{linear spin wave}
\newacronym{TFIM}{TFIM}{transverse field Ising model}
\newacronym{BKT}{BKT}{Berezinskii-Kosterlitz-Thouless}
\newacronym{SF}{SF}{spin-flip}
\newacronym{NSF}{NSF}{non-spin-flip}
\newacronym{INS}{INS}{inelastic neutron scattering}
\newacronym{QMC}{QMC}{quantum Monte-Carlo}

\begin{document}

\title{Field-Tuned Quantum Effects in a Triangular-Lattice Ising Magnet}

\author{Yayuan Qin}
\affiliation{State Key Laboratory of Surface Physics and Department of Physics, Fudan University, Shanghai 200433, China}

\author{Yao Shen}
\affiliation{State Key Laboratory of Surface Physics and Department of Physics, Fudan University, Shanghai 200433, China}
\affiliation{Condensed Matter Physics and Materials Science Department, Brookhaven National Laboratory, Upton, New York 11973, USA}

\author{Changle Liu}
\affiliation{State Key Laboratory of Surface Physics and Department of Physics, Fudan University, Shanghai 200433, China}

\author{Hongliang Wo}
\affiliation{State Key Laboratory of Surface Physics and Department of Physics, Fudan University, Shanghai 200433, China}
\affiliation{Shanghai Qizhi Institute, Shanghai 200232, China}

\author{Yonghao Gao}
\affiliation{State Key Laboratory of Surface Physics and Department of Physics, Fudan University, Shanghai 200433, China}

\author{Yu Feng}
\affiliation{Institute of High Energy Physics, Chinese Academy of Sciences (CAS), Beijing 100049, China}
\affiliation{Spallation Neutron Source Science Center, Dongguan 523803, China}
\affiliation{Materials and Life Science Division, J-PARC Center, Tokai, Ibaraki 319-1195, Japan}

\author{Xiaowen Zhang}
\affiliation{State Key Laboratory of Surface Physics and Department of Physics, Fudan University, Shanghai 200433, China}

\author{Gaofeng Ding}
\author{Yiqing Gu}
\affiliation{State Key Laboratory of Surface Physics and Department of Physics, Fudan University, Shanghai 200433, China}
\affiliation{Shanghai Qizhi Institute, Shanghai 200232, China}

\author{Qisi Wang}
\affiliation{State Key Laboratory of Surface Physics and Department of Physics, Fudan University, Shanghai 200433, China}
\affiliation{Physik-Institut, Universit{\"a}t Z{\"u}rich, Z{\"u}rich CH-8057, Switzerland}

\author{Shoudong Shen}
\affiliation{State Key Laboratory of Surface Physics and Department of Physics, Fudan University, Shanghai 200433, China}

\author{Helen C. Walker}
\author{Robert Bewley}
\affiliation{ISIS Facility, Rutherford Appleton Laboratory, STFC, Chilton, Didcot, Oxon OX11 0QX, United Kingdom}

\author{Jianhui Xu}
\thanks{Current address: Heinz Maier-Leibnitz Zentrum (MLZ), Technische Universit{\"a}t M{\"u}nchen, Garching 85748, Germany}
\affiliation{Helmholtz-Zentrum Berlin f{\"u}r Materialien und Energie GmbH, Berlin D-14109, Germany}

\author{Martin Boehm}
\author{Paul Steffens}
\affiliation{Institut Laue-Langevin, 38042 Grenoble Cedex 9, France}

\author{Seiko Ohira-Kawamura}
\author{Naoki Murai}
\affiliation{Materials and Life Science Division, J-PARC Center, Tokai, Ibaraki 319-1195, Japan}

\author{Astrid Schneidewind}
\affiliation{J{\"u}lich Centre for Neutron Science (JCNS) at Heinz Maier-Leibnitz Zentrum (MLZ), Forschungszentrum J{\"u}lich GmbH, Garching 85748, Germany}

\author{Xin Tong}
\affiliation{Institute of High Energy Physics, Chinese Academy of Sciences (CAS), Beijing 100049, China}
\affiliation{Spallation Neutron Source Science Center, Dongguan 523803, China}

\author{Gang Chen}\email[]{gangchen.physics@gmail.com}
\affiliation{Department of Physics and HKU-UCAS Joint Institute for Theoretical and Computational Physics at Hong Kong, The University of Hong Kong, Hong Kong, China}
\affiliation{State Key Laboratory of Surface Physics and Department of Physics, Fudan University, Shanghai 200433, China}
\affiliation{Collaborative Innovation Center of Advanced Microstructures, Nanjing University, Nanjing, 210093, China}

\author{Jun Zhao}\email[]{zhaoj@fudan.edu.cn}
\affiliation{State Key Laboratory of Surface Physics and Department of Physics, Fudan University, Shanghai 200433, China}
\affiliation{Shanghai Qizhi Institute, Shanghai 200232, China}
\affiliation{Institute of Nanoelectronics and Quantum Computing, Fudan University, Shanghai 200433, China}
\affiliation{Shanghai Research Center for Quantum Sciences, Shanghai 201315, China}
\affiliation{Collaborative Innovation Center of Advanced Microstructures, Nanjing University, Nanjing, 210093, China}

\maketitle

\textbf{We report thermodynamic and neutron scattering measurements of the triangular-lattice quantum Ising magnet {\TMGO} in longitudinal magnetic fields. Our experiments reveal a quasi-plateau state induced by quantum fluctuations. This state exhibits an unconventional non-monotonic field and temperature dependence of the magnetic order and excitation gap. In the high field regime where the quantum fluctuations are largely suppressed, we observed a disordered state with coherent magnon-like excitations despite the suppression of the spin excitation intensity. Through detailed semi-classical calculations, we are able to understand these behaviors quantitatively from the subtle competition between quantum fluctuations and frustrated Ising interactions.}

\textbf{keyword:} Neutron scattering, Quantum magnet, Magnetic frustration, Transverse field Ising model

\section{1. Introduction}

Ising model is a stereotype model in modern statistical physics and has revolutionarily advanced our understanding of phase transitions \cite{Onsager}. It can be realized in quantum magnets with a strong easy-axis anisotropy that pins the spin moments towards a fixed axis. Due to the pinning effect, the quantum effect is suppressed, and the physics becomes rather classical \cite{Ising_Ramirez}. To introduce quantum effects, one usually applies an {\sl external} magnetic field along the transverse direction and creates a quantum Ising model \cite{Quantum_Sachdev, TranField_Moessner, TranField_Savary, Mikeska1991, Bitko1996}. Representative examples of this type include the quasi-one-dimensional magnets CoNb$_2$O$_6$ (Ref.~\onlinecite{CoNb2O6_Coldea}), BaCo$_2$V$_2$O$_8$ (Refs.~\onlinecite{BCVO_Faure, BCVO_Matsuda, BCVO_Zou}), and SrCo$_2$V$_2$O$_8$ (Refs.~\onlinecite{SCVO_Wang1, SCVO_Wang2}), in which quantum criticality and novel transitions have been reported. Nevertheless, the experimental realization of quantum Ising model in two dimensional systems is rare despite decades of theoretical efforts \cite{TranField_Moessner, TFIM_Moessner, TFIM_Isakov, TFIM_Damle, TFIM_Biswas, TFIM_Wang}. In contrast to this conventional route, nature provides a distinct example of quantum Ising magnets that build quantum mechanics intrinsically in the system. The observation is that, although the exchange interaction is primarily Ising like, the intrinsic {\gls*{CEF}} splitting of the Ising moment naturally creates quantum effects out of these classically interacting degrees of freedom \cite{TranField_Wang, TranField_Gang, TranField_Dun}. These systems are quoted as ``intrinsic quantum Ising magnets''.

The triangular-lattice antiferromagnet {\TMGO} (ref. \onlinecite{TMGO_Cava}) is a promising candidate for such requirements. Due to the strong {\gls*{SOC}} and {\gls*{CEF}} splitting, the {\gls*{CEF}} ground state wave function of Tm$^{3+}$ ions is dominated by $J^z = \pm 6$, leading to a large magnetic moment of 6.59~$\mu_\mathrm{B}$/Tm$^{3+}$ and Ising spin nature \cite{TMGO_Yao}. Thus, quantum effects are expected to be significantly suppressed. Interestingly, however, the {\TMGO} {\gls*{CEF}} ground state is a quasi-doublet composed of two singlets separated by a small energy gap that can be mapped into an effective transverse field $h_y$ (Fig.~\ref{fig:diagram}a). In this case, the transverse field, which is intrinsic in origin and homogeneous in general, leads to quantum tunneling effects among various Ising spin configurations and strongly compete with the Ising-type interactions \cite{TranField_Wang, TranField_Gang, TranField_Dun}. Different from the coherent quantum fluctuations in quantum spin liquids that lead to long-range entanglement \cite{QSL_Balents, QSL_Zhou, QSL_Yao1, QSL_Yao2}, the quantum effects here are of single-ion level and will potentially stabilize three-sublattice ordering. Meanwhile, short-range quantum entanglement emerges from the non-commutativeness between transverse field and Ising interactions, which may renormalize the magnetic interactions. The magnetic properties can be effectively described by the {\gls*{TFIM}} \cite{TMGO_CLiu, TMGO_Yao}
\begin{align}
\mathcal{H} &= \sum_{ij}J^{zz}_{ij}S^z_i S^z_j-\sum_{i}(h_yS^y_i+B_zS^z_i),
\end{align}
where the Ising exchange interactions can be kept within the first few neighbors and are geometrically frustrated. Intriguingly, due to the large $g$-factor in {\TMGO} ($g_{\parallel} \sim$ 13.18) \cite{TMGO_YLi1}, the contribution from external longitudinal field, $B_z = g_{\parallel}\mu_\mathrm{B} B$, is comparable to the Ising interactions and quantum fluctuations so that the interplay between quantum and classical contributions can be easily tuned by external field. Therefore, {\TMGO} provides a unique platform to manipulate the quantum effects in a controlled manner.

Moreover, in rare-earth materials, due to the complex {\gls*{SOC}} and {\gls*{CEF}} splitting, the pseudo-spins can host multipolar behaviors \cite{Multipole_Santini, Multipole_Alistair, Multipole_CLiu}. In the case of {\TMGO}, we find that the transverse components of the pseudo-spins, $S^x$ and $S^y$, behave as multipoles that cannot be directly detected by neutron diffraction while the longitudinal one, $S^z$, remains dipolar. Its zero-field magnetic ground state is an intertwined dipolar and multipolar order in which the dipolar $S^z$ forms a three-sublattice clock phase and the multipolar $S^x/S^y$ components are ferro-aligned \cite{TMGO_CLiu, TMGO_Yao}. The observed spin excitations are in a reasonable agreement with the {\gls*{LSW}} theory, in which, however, only $S_{zz}$ channel is detectable, rising from the coherent spin wave excitations and fluctuations of the multipolar components \cite{Multipole_CLiu, Multipole_Kadowaki}. In this paper, we continue this research and study the evolution of intrinsic quantum properties of {\TMGO} in longitudinal fields.

\section{2. Results}

\subsection{2.1 Heat capacity and neutron diffraction}

\begin{figure}[t]
\includegraphics{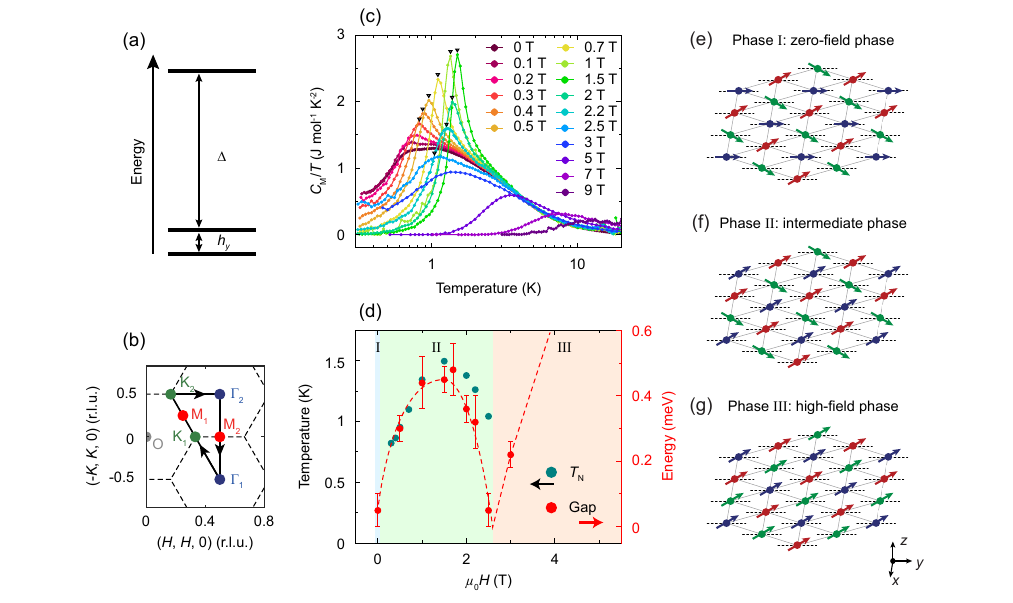}
\caption{Heat capacity and magnetic phase diagram of {\TMGO}. (a) Schematic of the low energy {\gls*{CEF}} levels of Tm$^{3+}$ ions in {\TMGO}. The ground-state quasi-doublet is well separated from the high-energy excited state ($\Delta \sim$ 38 meV), indicating an effective spin-1/2 state at low temperature \cite{TMGO_YLi1, TMGO_Yao}. (b) Sketch of the reciprocal space in $xy$ plane. The dashed lines denote Brillouin zone boundaries and the wavevector is defined as $\textbf{Q}=H\textbf{a}^*+K\textbf{b}^*+L\textbf{c}^*$; r.l.u., reciprocal lattice unit. (c) Magnetic heat capacity in various longitudinal fields ($H \parallel z$). The phonon contributions were subtracted by measuring the non-magnetic iso-structural LuMgGaO$_4$ single crystal. The transition temperatures are marked by black triangles. (d) Magnetic phase diagram consisting of field dependent {\TN} determined from heat capacity, and spin gaps at K points determined from INS data below 0.15~K. The phase diagram can be divided into three parts. \textit{Phase I}: zero-field phase; \textit{Phase II}: intermediate phase; \textit{Phase III}: high-field phase. The dashed line is a guide to the eye. (e)-(g) Schematics of the semi-classical spin structures regarding the three phases in (d). The dashed lines denote the virtual $y$ direction and different colors indicate three different sublattices.}
\label{fig:diagram}
\end{figure}

\begin{figure}[t]
\includegraphics{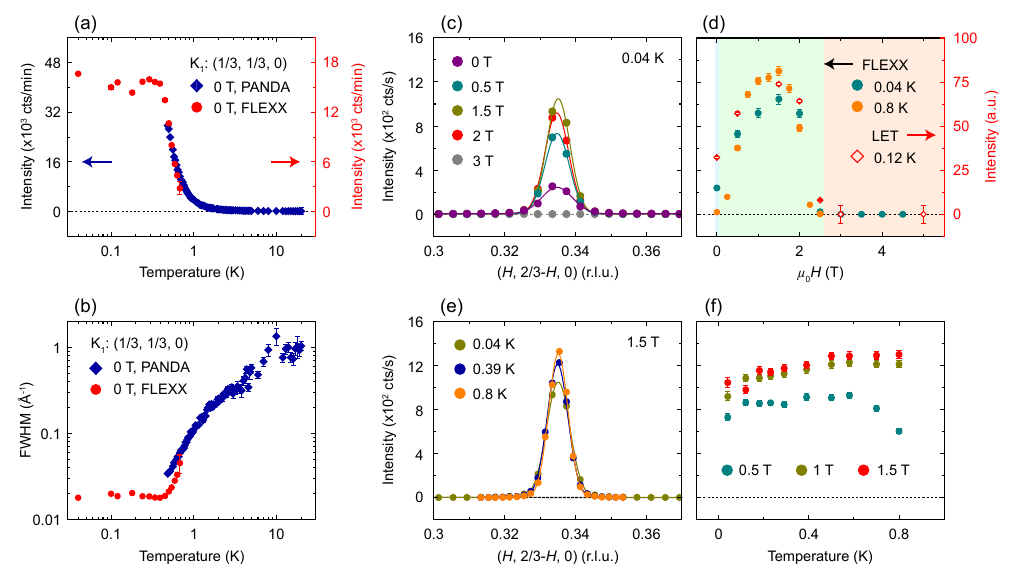}
\caption{Neutron diffraction data as a function of field and temperature. (a),(b) Temperature dependence of the peak intensities and full-width at half-maximum (FWHM) of the magnetic Bragg peak at K$_1$ point. (c) Qscans near the magnetic peak, {\Q} = (1/3, 1/3, 0), in different longitudinal fields. (d) Field dependence of the fitted amplitude of the magnetic peaks at different temperatures. (e) Representative Qscans near the magnetic peak at different temperatures in external field of 1.5~T. (f) Temperature dependence of the fitted peak amplitude at {\Q} = (1/3, 1/3, 0) in different fields. The solid lines in (c) and (e) are the fitting results with Gaussian profiles and a flat background, which are also used to evaluate the peak intensities and FWHM in (a), (b), (d) and (f). Data in (c), (e) and (f) are collected at FLEXX spectrometer. a.u., arbitrary unit; cts/s, counts per second; cts/min, counts per minute; error bars, 1 s.d.}
\label{fig:bragg}
\end{figure}

We start by reviewing the phase transitions of {\TMGO} in absence of external field. Although no $\lambda$-shaped transition is present in the heat capacity data (Fig.~\ref{fig:diagram}c), a well-defined magnetic Bragg peak associated with the three-sublattice spin order is observed at the K point, {\Q} = (1/3, 1/3, 0), at low temperature (Fig.~\ref{fig:bragg}c). With temperature decreasing from 20 K, the intensity gradually increases with reduced peak width (Fig.~\ref{fig:bragg}a,~b). The most abrupt change takes place around 1~K, which corresponds to a shoulder-like anomaly in the heat capacity (Fig.~\ref{fig:diagram}c). Both the peak intensity and peak width saturate below $\sim$0.4~K. When a small magnetic field is applied along the longitudinal $z$ direction, the heat capacity shows a sharp anomaly, indicating a new magnetic phase transition (Fig.~\ref{fig:diagram}c). The transition temperature, {\TN}, increases with increasing field, reaches a maximum at $\sim$1.5~T, and then decreases with further increasing field (Fig.~\ref{fig:diagram}d). The transition completely vanishes above $\sim$2.7~T.

To elucidate the nature of the phase transitions in longitudinal fields, we performed neutron diffraction measurements at various temperatures. As shown in Fig.~\ref{fig:bragg}c,~d, the magnetic Bragg peak intensity at K point with temperatures of 0.04, 0.12 and 0.8~K also shows a maximum around 1.5~T, and disappears at $\sim$2.7~T. This indicates that the phase transition seen in heat capacity measurements is also associated with the K point order. Moreover, the peak intensities in magnetic fields of 0.5, 1, and 1.5~T are enhanced in the intermediate temperature range (0.2 $\lesssim T \lesssim$ 0.8 K) below {\TN}, distinct from a conventional order parameter behavior (Fig.~\ref{fig:bragg}e,~f). The unusual enhancement of the magnetic peak on warming is due to the competition between thermal and quantum fluctuations, as will be discussed subsequently.

\subsection{2.2 Inelastic neutron scattering}

\begin{figure}[t]
\includegraphics{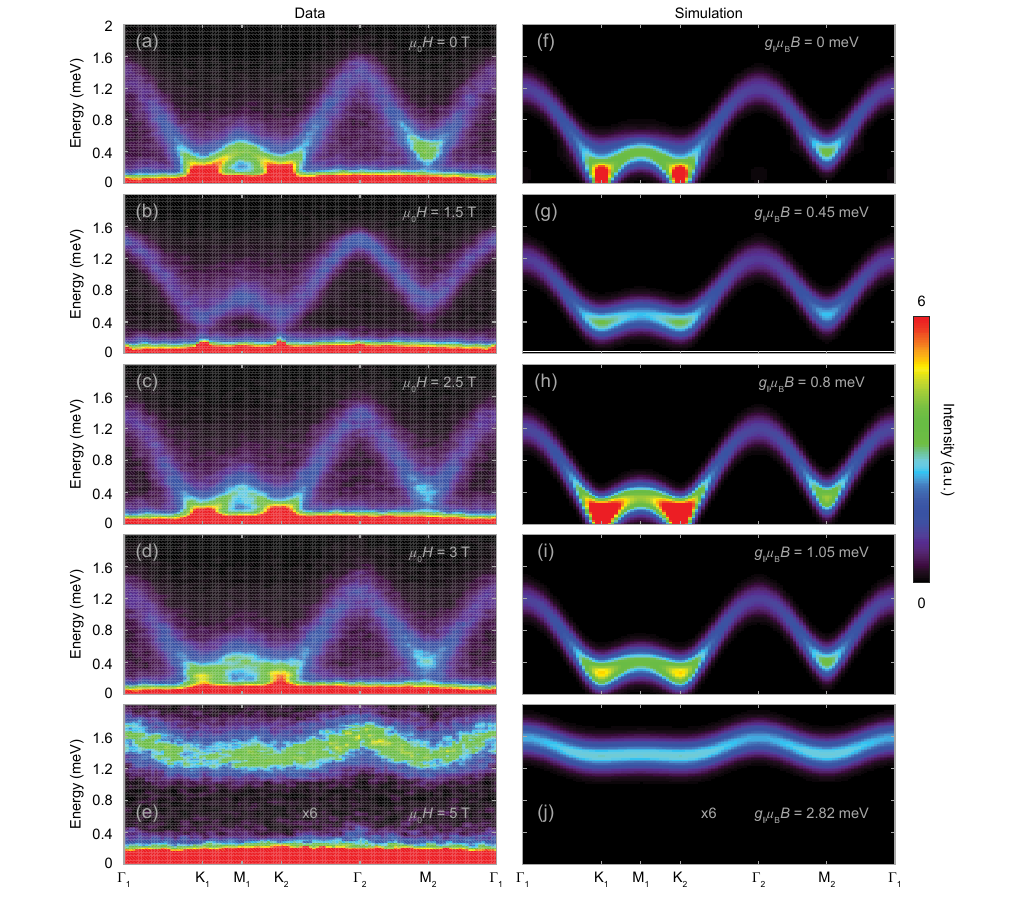}
\caption{Spin excitations in different longitudinal fields at 0.12~K. (a)-(e) INS spectra of the spin excitations in {\TMGO} along the high symmetry directions marked by the black arrows in Fig.~1b in the indicated fields. (f)-(j) Calculated spin wave dispersions using the model specified in the main text. Intensities in (e) and (j) are multiplied by 6. The intensity scale for the inelastic spectra is uniform throughout the text.}
\label{fig:disp}
\end{figure}

The enhancement of magnetic order in longitudinal fields has been interpreted either as the consequence of quantum fluctuations in {\gls*{TFIM}} \cite{TMGO_CLiu}, or large {\gls*{CEF}} randomness induced by structural disorder \cite{TMGO_YLi1}. To further distinguish these two scenarios, we used {\gls*{INS}} to measure the detailed field dependence of the spin excitations in {\TMGO}. Figure \ref{fig:disp}a-e exhibit a series of neutron spectra along the high symmetry directions in various fields. Compared with the zero-field data in Fig.~\ref{fig:disp}a, the overall spin excitations in external field of 1.5~T tend to shift towards higher energies, leading to an enlarged spin gap (Fig.~\ref{fig:disp}b). This is clearly inconsistent with the {\gls*{CEF}} randomness picture in which it was suggested that the system would be gapless in external field \cite{TMGO_YLi1}. Instead, it can be readily understood without introducing much structural disorder. Below 1.5~T, the spin gap increases with increasing field (Fig.~\ref{fig:diagram}d). Thus, the system needs more energy to excite the quasi-particles from the magnetic ground state, implying a more stable order. Consequently, the magnetic Bragg peaks at K points are getting stronger and {\TN} is higher. Above 1.5~T, the spin gap gradually decreases with field and both the magnetic peak intensity and {\TN} are reduced (Fig.~\ref{fig:diagram}d,~\ref{fig:bragg}d). When the external field goes across the transition of $\sim$2.7~T into the high-field regime, the spin gap reappears, indicating that the external field overcomes the spin-spin correlations and the Zeeman term dominates the spin dynamics, making the excitations less dispersive (Fig.~\ref{fig:disp}e). The relatively sharp excitations observed in various fields suggest that the structural disorder is not significant in {\TMGO}.

The Ising interactions arise from the exchange and dipole-dipole interactions among Tm$^{3+}$ local moments. For a classical Ising model, the excitations correspond to Ising spin flipping and are non-dispersive. In {\TMGO}, due to the presence of intrinsic transverse field, the excitations become dispersive. To further determine the nature of the spin excitations, we carried out polarized neutron scattering measurements on {\TMGO}. It is shown that the spin excitations are essentially longitudinally polarized in the $S_{zz}$ channel while the transverse $S_{+-}$ component is absent (Fig.~\ref{fig:Hdep}a) (Supplementary material). This is completely different from the classical spin wave with only transverse excitations at low energies. Therefore, the longitudinal channel of {\gls*{INS}} spectra, $S_{zz}$, manifests the quantumness of the excitations that are strictly forbidden for classical spins.

\subsection{2.3 Phase diagram}

\begin{figure}[t]
\includegraphics{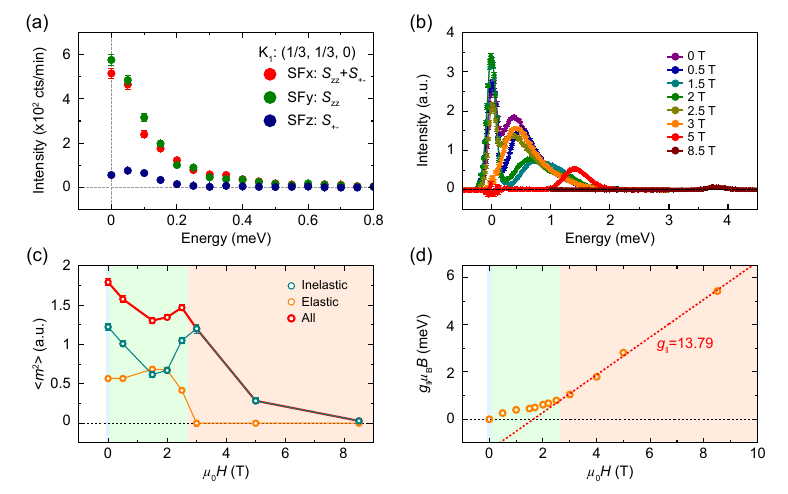}
\caption{Polarized neutron scattering data and the field dependence of the spin excitations. (a) Energy scans at K$_1$, 0~T and 1.7~K in all the three spin-flip (SF) channels which correspond to different excitation components (Supplementary material). The tiny peak in SFz channel may come from the leakage of non-spin-flip channel. (b) Energy dependence of the scattering function integrated through the whole Brillouin zone (-0.5 $< L <$ 0.5) in different external fields at 0.12~K. (c) Field dependence of the total local moment $\langle m^2\rangle$ (\textit{All}) and the separated \textit{elastic} and \textit{inelastic} contributions (Supplementary material). (d) The external field term, $g_{\parallel}\mu_\mathrm{B} B$, used during the {\gls*{LSW}} calculation versus the experimentally applied field values, $\mu_0 H$. The dashed red line denotes the linear fitting of the data points in Phase III, which gives the $g$-factor along the $z$ direction.}
\label{fig:Hdep}
\end{figure}

After confirming the quantum fluctuations in {\TMGO}, we can understand the phase diagram comprehensively. Due to the single-ion quantum fluctuations, the dressed Ising moments are no longer homogeneous but modulated from site to site, the order of which resembles the spin density wave in quantum spin chains \cite{BCVO_Grenier}. In zero field, the competition between classical Ising interactions and quantum fluctuations drives the system into a three-sublattice phase with up-down-zero configuration in each triangle (Phase I in Fig.~\ref{fig:diagram}d). The quantum fluctuation in this phase is so strong that the magnetic moments are significantly reduced. In weak longitudinal fields, the classical triangular-lattice Ising magnet would evolve into a two-up-one-down 1/3-plateau state \cite{Schick1977, Miyashita1986, Honecker1999, XXZ_Sellmann, XXZ_Yamamoto}. In {\TMGO}, the quantum fluctuations turn it into a quasi-plateau phase with modulated spin moments (Phase II). Here, the external field acts as an extra classical contribution and suppresses the quantum effects. As a result, the magnetic order is enhanced with stronger magnetic peaks at K points, higher {\TN} and a larger spin gap. Since the spin dynamics here are of quantum origin, the {\gls*{INS}} intensity is weaker (Fig.~\ref{fig:Hdep}b, c). Moreover, at higher temperatures, the quantum effects are partially quenched due to the thermal fluctuations, leading to enhanced magnetic Bragg peaks at higher temperatures (Fig.~\ref{fig:bragg}e,~f). When the field is further increased towards the transition field $\mu_0 H_\mathrm{c} \sim$ 2.7~T, the quantum fluctuation is enhanced again and the rebuilt quantum effects suppress the static magnetic order, reduce the spin gap and lower the transition temperature while the quantum excitations grow in intensity (Fig.~\ref{fig:diagram}d, Fig.~\ref{fig:Hdep}b, c). Above $\mu_0 H_\mathrm{c}$, the three-sublattice magnetic order is completely eliminated and the system enters the nearly polarized state (Phase III), making the magnetic state more classical. Meanwhile, the spin gap increases with field and the quantum excitations are progressively suppressed (Fig.~\ref{fig:disp}c-e).

\subsection{2.4 Spin wave calculation}

To quantitatively describe the field tuned quantum effects, we utilize mean-field approach to solve {\gls*{TFIM}} by introducing the virtual $y$ axis, which lies in the structural $ab$ plane but does not correspond to any real direction, and can even vary from site to site. In this semi-classical picture, the pseudo-spin has a uniform size but becomes tilted from the Ising axis to $y$ direction owing to the transverse field. The effective spin structures of the three phases are illustrated in Fig.~\ref{fig:diagram}e-g. Since the transverse components of the effective spins, $S^x$ and $S^y$, transform as multipoles, only the moment projection in $z$ direction contributes to the magnetic Bragg peak, making the peak intensity at K point varies with longitudinal field (Fig.~\ref{fig:bragg}d). Regarding the {\gls*{INS}} process, measuring $S^z$ moment will flip the multipolar components and trigger coherence spin wave excitations in $S_{zz}$ channel \cite{Multipole_CLiu, pyrochlore_YLi}, consistent with our polarized neutron measurements. Based on this, we use {\gls*{LSW}} theory through the \textsc{spinw} program to simulate the quantum excitations of {\gls*{TFIM}} in longitudinal fields \cite{SPINW}. The Ising interactions and transverse field term are determined by fitting the zero-field spin excitation spectra, which are $J^{zz}_1$ = 0.54 meV, $J^{zz}_2$ = 0.026 meV, $h_y$ = 0.62 meV \cite{TMGO_Yao}. We show that the {\gls*{LSW}} theory can describe the observed quantum excitations throughout the phase diagram (Fig.~\ref{fig:disp}f-j). The field-dependent evolution of the magnetic Bragg peak, spin gap, overall spectra intensity and dispersions are reproduced reasonably well by the {\gls*{LSW}} theory (Supplementary material).We note that the simulated magnetic field is slightly smaller than the applied field (Fig.~\ref{fig:Hdep}d); this is probably due to the collective quantum effect that is not considered in the {\gls*{LSW}} calculations. More sophisticated calculation such as the renormalized spin wave theory that takes into account the quantum correction to ordered moments may further refine this analysis. A recent {\gls*{QMC}} calculations actually suggested a slightly larger exchange coupling constants than the {\gls*{LSW}} theory calculations, but the ratio of $J_2/J_1$ remains nearly the same, which is not surprising and does not qualitatively change the magnetic structure and phase diagram based on semi-classical calculations \cite{TMGO_WLi,TMGO_CLiu}. When the external field is relatively large, collective quantum effects are largely suppressed and the {\gls*{LSW}} calculation becomes more accurate. We can extract the effective $g$-factor of 13.79 from the high field fitting (Fig.~\ref{fig:Hdep}d), which is close to the reported value determined by magnetization measurement \cite{TMGO_YLi1}.

\section{3. Discussion and conclusion}

It has also been suggested theoretically that Phase I will melt in a two-step manner through two {\gls*{BKT}} transitions and the intermediate {\gls*{BKT}} phase hosts a emergent U(1) symmetry \cite{TFIM_Isakov, TFIM_Damle, TFIM_Biswas, TMGO_CLiu, TMGO_WLi}. Furthermore, it is predicted that in the {\gls*{BKT}} phase the magnetic susceptibility will diverge in the small longitudinal field limit with a unique scaling behavior \cite{TFIM_Damle, TFIM_Biswas}. However, this is not observed in our magnetization measurements (Supplementary material). A recent work reports susceptibility data and numerical simulation showing {\gls*{BKT}} behavior between 0.6 and 0.9~T \cite{NMR_BKT} which, however, lies deep into Phase II of our phase diagram. It has been well established that as a quasi-long-range ordered state, the {\gls*{BKT}} phase is fragile to perturbations and can be easily killed by external field ($<$ 0.1~T) as indicated by {\gls*{QMC}} calculations \cite{QMC_Meng}. Thus, the claimed evidence for {\gls*{BKT}} physics for the up-down-zero phase and the comparison are then irrelevant. In addition, no divergence of susceptibility was found in ref.~\onlinecite{NMR_BKT} near the zero-field regime. It should be noted that according to {\gls*{QMC}} simulation \cite{QMC_Meng}, the upper {\gls*{BKT}} transition would involve into a second-order transition in longitudinal field, which is also inconsistent with ref.~\onlinecite{NMR_BKT}. An earlier calculation suggested that a new magnetic peak at M point should appear at high temperatures owing to proliferated vortex-antivortex pairs \cite{TMGO_WLi}, which is not observed in our neutron scattering measurements (Supplementary material).

To summarize, we have performed neutron scattering measurements on the intrinsic quantum Ising magnet {\TMGO} in longitudinal fields. In weak field, the zero-field three-sublattice order is replaced by a intermediate quasi-plateau phase, in which both the static moments and spin gap behave in a non-monotonic manner as a function of external field. In high fields, the system is driven into the field-induced polarized state. Through a semi-classical analysis, we show that the observed magnetic order and the associated quantum excitations can be well described by {\gls*{TFIM}} throughout the phase diagram with the subtle competition among quantum fluctuations, frustrated Ising interactions, and thermal fluctuations. Our results demonstrate that {\TMGO}, as a rare experimental realization of 2D {\gls*{TFIM}}, is a remarkable platform to illustrate the interplay between quantum and classical interactions, and that the associated quantum states can be manipulated through external fields in a highly controlled manner.

\section{Declaration of Competing Interest}

The authors declare that they have no conflict of interest.

\section{Acknowledgments}

This work was supported by the Innovation Program of Shanghai Municipal Education Commission (grant no. 2017-01-07-00-07-E00018), the National Key R\&D Program of the MOST of China (grant nos. 2016YFA0300203, 2016YFA0300500, 2016YFA0301001, 2018YFE0103200), the National Natural Science Foundation of China (grant No. 11874119), Shanghai Municipal Science and Technology Major Project with Grant No.2019SHZDZX04, and the Hong Kong Research Grants Council (GRF no. 17303819, 17306520). Y.F. and X.T. was supported by the National Nature Science Foundation of China (No. 11875265), the Scientific Instrument Developing Project of the Chinese Academy of Sciences (3He based neutron polarization devices), the Institute of High Energy Physics, and the Chinese Academy of Science. The datasets for the inelastic neutron scattering experiment on the time-of-flight LET spectrometer are available from the ISIS facility, Rutherford Appleton Laboratory data portal (10.5286/ISIS.E.RB1910127). The datasets for the polarized neutron scattering experiment on the cold triple-axis ThALES spectrometer are available from the Institute Laue-Langevin data portal (https://doi.ill.fr/10.5291/ILL-DATA.4-05-755). The neutron experiment at the Materials and Life Science Experimental Facility of the J-PARC was performed under a user program (Proposal No. 2019A0116). All other data that support the plots within this paper and other findings of this study are available from the corresponding authors upon reasonable request.

\section{Author contributions}

Jun Zhao and Gang Chen planned the project. Yayuan Qin, Yao Shen, and Shoudong Shen synthesized the sample. Yao Shen and Xiaowen Zhang characterized the sample. Yayuan Qin, Yao Shen, Hongliang Wo, Yu Feng, Gaofeng Ding, Yiqing Gu, Qisi Wang, Helen C. Walker, Xin Tong and Jun Zhao carried out the neutron experiments with experimental assistance from Robert Bewley, Jianhui Xu, Martin Boehm, Paul Steffens, Seiko Ohira-Kawamura, Naoki Murai, and Astrid Schneidewind. Jun Zhao, Yayuan Qin and Yao Shen analyzed the data. Changle Liu, Yonghao Gao, and Gang Chen provided the theoretical support and analysis. Jun Zhao, Gang Chen, Yao Shen, Changle Liu, and Yonghao Gao wrote the paper. All authors provided comments on the paper. Yayuan Qin and Yao Shen contributed equally to this work.

\newpage

\begin{thebibliography}{}

\bibitem{Onsager} Onsager L. Crystal statistics. I. A two-dimensional model with an order-disorder transition. Phys Rev 1994;65:117.

\bibitem{Ising_Ramirez} Ramirez A. Strongly geometrically frustrated magnets. Annu Rev Mater Sci 1994;24:453.

\bibitem{Quantum_Sachdev} Sachdev S. Quantum phase transitions (Cambridge University Press, Cambridge, England, 2011).

\bibitem{TranField_Moessner} Moessner R, Sondhi S L, Chandra P. Two-dimensional periodic frustrated Ising models in a transverse field. Phys Rev Lett 2000;84:4457.

\bibitem{TranField_Savary} Savary L, Balents L. Disorder-induced quantum spin liquid in spin ice pyrochlores. Phys Rev Lett 2017;118:087203.

\bibitem{Mikeska1991} Mikeska H-J, Steiner M. Solitary excitations in one-dimensional magnets. Adv Phys 1991:40:191.

\bibitem{Bitko1996} Bitko D, Rosenbaum T F, Aeppli G. Quantum critical behavior for a model magnet. Phys Rev Lett 1996;77:940.

\bibitem{CoNb2O6_Coldea} Coldea R, Tennant D A, Wheeler E M, et al. Quantum criticality in an Ising chain: Experimental evidence for emergent $E_8$ symmetry. Science 2010;327:177.

\bibitem{BCVO_Faure} Faure Q, Takayoshi S, Petit S, et al. Topological quantum phase transition in the Ising-like antiferromagnetic spin chain BaCo$_2$V$_2$O$_8$. Nat Phys 2018;14:716.

\bibitem{BCVO_Matsuda} Matsuda M, Onishi H, Okutani A, et al. Magnetic structure and dispersion relation of the $S$=1/2 quasi-one-dimensional Ising-like antiferromagnet BaCo$_2$V$_2$O$_8$ in a transverse magnetic field. Phy Rev B 2017;96:024439.

\bibitem{BCVO_Zou} Zou H, Cui Y, Wang X, et al. Exceptional E$_8$ symmetry in spin dynamics of quasi-one-dimensional antiferromagnet BaCo$_2$V$_2$O$_8$. arXiv: 2005.13302 (2020).

\bibitem{SCVO_Wang1} Wang Z, Wu J, Yang W, et al. Experimental observation of Bethe strings. Nature 2018;554:219.

\bibitem{SCVO_Wang2} Wang Z, Lorenz T, Gorbunov D I, et al. Quantum criticality of an Ising-like spin-1/2 antiferromagnetic chain in a transverse magnetic field. Phys Rev Lett 2018;120:207205.

\bibitem{TFIM_Moessner} Moessner R, Sondhi S L. Ising models of quantum frustration. Phys Rev B 2001;63:224401.

\bibitem{TFIM_Isakov} Isakov S V, Moessner R. Interplay of quantum and thermal fluctuations in a frustrated magnet. Phys Rev B 2003;68:104409.

\bibitem{TFIM_Damle} Damle K. Melting of three-sublattice order in easy-axis antiferromagnets on triangular and Kagome lattices. Phys Rev Lett 2015;115:127204.

\bibitem{TFIM_Biswas} Biswas S, Damle K. Singular ferromagnetic susceptibility of the transverse-field Ising antiferromagnet on the triangular lattice. Phys Rev B 2018;97:085114.

\bibitem{TFIM_Wang} Wang Y, Humeniuk S, Wan Y. Tuning the two-step melting of magnetic order in a dipolar Kagome spin ice by quantum fluctuations. Phys Rev B 2020;101:134414.

\bibitem{TranField_Wang} Wang Y, Cooper B R. Collective excitations and magnetic ordering in materials with singlet crystal-field ground state. Phys Rev 1968;172:539-551.

\bibitem{TranField_Gang} Chen G. Intrinsic transverse field in frustrated quantum Ising magnets: Physical origin and quantum effects. Phys Rev Res 2019;1:033141.

\bibitem{TranField_Dun} Dun Z, Bai X, Paddison J A M, et al. Quantum versus classical spin fragmentation in dipolar Kagome ice Ho$_3$Mg$_2$Sb$_3$O$_{14}$. Phys Rev X 2020;10:031069.

\bibitem{TMGO_Cava} Cevallos A F, Stolze K, Kong T, et al. Anisotropic magnetic properties of the triangular plane lattice material TmMgGaO$_4$. Mater Res Bull 2018;105:154-158.

\bibitem{TMGO_Yao} Shen Y, Liu C, Qin Y, et al. Intertwined dipolar and multipolar order in the triangular-lattice magnet TmMgGaO$_4$. Nat Commun 2019;10:4530.

\bibitem{QSL_Balents} Balents L. Spin liquids in frustrated magnets. Nature 2010;464:199.

\bibitem{QSL_Zhou} Zhou Y, Kanoda K, Ng T-K. Quantum spin liquid states. Rev Mod Phys 2017;89:025003.

\bibitem{QSL_Yao1} Shen Y, Li Y-D, Wo H, et al. Evidence for a spinon Fermi surface in a triangular-lattice quantum-spin-liquid candidate. Nature 2016;540:559.

\bibitem{QSL_Yao2} Shen T, Li Y-D, Walker H C, et al. Fractionalized excitations in the partially magnetized spin liquid candidate YbMgGaO$_4$. Nat Commun 2018;9:4138.

\bibitem{TMGO_CLiu} Liu C, Huang C-J, Chen G. Intrinsic quantum Ising model on a triangular lattice magnet TmMgGaO$_4$. Phys Rev Research 2020;2:043013.

\bibitem{TMGO_YLi1} Li Y, Bachus S, Deng H, et al. Partial up-up-down order with the continuously distributed order parameter in the triangular antiferromagnet TmMgGaO$_4$. Phys Rev X 2020;10:011007.

\bibitem{Multipole_Santini} Santini P, Carretta S, Amoretti G, et al. Multipolar interactions in $f$-electron systems: the paradigm of actinide dioxides. Rev Mod Phys 2009;81:807-863.

\bibitem{Multipole_Alistair} Alistair C S, Gerd F, Dmytro I S. Multipolar phases and magnetically hidden order: Review of the heavy-fermion compound Ce$_{1-x}$La$_x$B$_6$. Rep Prog Phys 2016;79:066502.

\bibitem{Multipole_CLiu} Liu C, Li Y, Chen G. Selective measurements of intertwined multipolar orders: Non-Kramers doublets on a triangular lattice. Phys Rev B 2018;98:045119.

\bibitem{Multipole_Kadowaki} Kadowaki H, Takatsu H, Taniguchi T, et al. Composite spin and quadrupole wave in the ordered phase of Tb$_{2+x}$Ti$_{2-x}$O$_{7+y}$. SPIN 2015;5:1540003.

\bibitem{BCVO_Grenier} Grenier B, Simonet V, Canals B, et al. Neutron diffraction investigation of the H-T phase diagram above the longitudinal incommensurate phase of BaCo$_{2}$V$_{2}$O$_{8}$. Phys Rev B 2015;92:134416.

\bibitem{Schick1977} Schick M, Walker J S, Wortis M. Phase diagram of the triangular Ising model: Renormalization-group calculation with application to adsorbed monolayers. Phys Rev B 1977;16:2205.

\bibitem{Miyashita1986} Miyashita S. Magnetic properties of Ising-like Heisenberg antiferromagnets on the triangular lattice. J Phys Soc Jpn 1986;55:3605.

\bibitem{Honecker1999} Honecker A. A comparative study of the magnetization process of two-dimensional antiferromagnets. J Phys: Condens Matter 1999;11:4697.

\bibitem{XXZ_Sellmann} Sellmann D, Zhang X, Eggert S. Phase diagram of the antiferromagnetic $\mathrm{XXZ}$ model on the triangular lattice. Phys Rev B 2015;91:081104.

\bibitem{XXZ_Yamamoto} Yamamoto D, Marmorini G, Danshita I. Quantum phase diagram of the triangular-lattice $\mathrm{XXZ}$ model in a magnetic field. Phys Rev Lett 2014;112:127203.

\bibitem{pyrochlore_YLi} Li Y, Chen G. Symmetry enriched U(1) topological orders for dipole-octupole doublets on a pyrochlore lattice. Phys Rev B 2017;95:041106.

\bibitem{SPINW} Toth S, Lake B. Linear spin wave theory for single-$\ensuremath{Q}$ incommensurate magnetic structures. J Phys: Condens Matter 2015;27:166002.

\bibitem{TMGO_WLi} Li H, Liao Y-D, Chen B, et al. Kosterlitz-Thouless melting of magnetic order in the triangular quantum Ising material TmMgGaO$_4$. Nat Commun 2020;11;1111.

\bibitem{NMR_BKT} Hu Z, Ma Z, Liao Y-D, et al. Evidence of the Berezinskii-Kosterlitz-Thouless phase in a frustrated magnet. Nat Commun 2020;11:5631.

\bibitem{QMC_Meng} Liao Y D, Li H, Yan z, et al. Phase diagram of the quantum Ising model on a triangular lattice under external field. Phys Rev B 2021;103:104416.

\end{thebibliography}
\end{document}

\title{Supplementary Material: Field-Tuned Quantum Effects in a Triangular-Lattice Ising Magnet}

\renewcommand{\thepage}{S\arabic{page}} 
\renewcommand{\thesection}{S\arabic{section}}  
\renewcommand{\thetable}{S\arabic{table}}  
\renewcommand{\thefigure}{S\arabic{figure}}

\maketitle

\section{Sample synthesis and Neutron scattering experiments.}

\begin{figure*}[bt]
\includegraphics{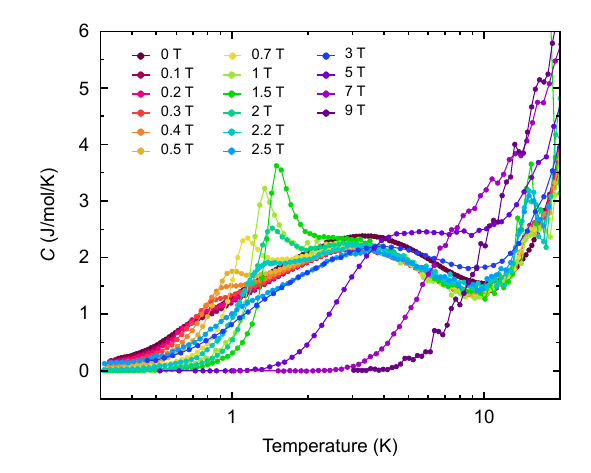}
\caption{Heat capacity in various longitudinal fields without phonon subtraction. The phonon subtracted data are presented in Fig. 1c.}
\label{fig:HC}
\end{figure*}

Large size {\TMGO} single crystals were synthesized using the floating-zone technique. The heat capacity, X-ray Laue and X-ray diffraction measurements indicate the high crystallization quality of the sample (Fig.~\ref{fig:HC}) \cite{TMGO_Yao}. The neutron diffraction measurements were carried out on the cold three axes spectrometer PANDA at the Heinz Maier-Leibnitz Zentrum, Garching, Germany \cite{PANDA}, and the FLEXX cold triple-axis spectrometer in the BER-II reactor at Helmholtz-Zentrum Berlin, Germany \cite{FLEXX}. For the PANDA experiment, a vertically focused PG(002) was used as a monochromator and analyzer. We fixed the final neutron energy at $E_\mathrm{f}$ = 4.06~meV, leading to an energy resolution of around 0.1~meV. In order to avoid the contamination from neutrons with higher orders, a Be filter was placed after the sample. One piece of single crystal with mass of 3 grams was aligned in the ($H$, $K$, 0) plane. A closed-cycle refrigerator equipped with a $^3$He insert was used to reach the base temperature of 0.485~K. For the FLEXX experiment, we used a PG(002) as both the monochromator and analyzer, and the final neutron energy was fixed at $E_\mathrm{f}$ = 3.5~meV. A velocity selector installed in front of the monochromator was used to remove the higher-order neutrons. One piece of single crystal (1.7~g) was aligned in the ($H$, $K$, 0) plane for the measurement. The sample was attached to a dilution insert and put into a vertical magnet to reach the base temperature of 40~mK.

\begin{figure*}[bt]
\includegraphics{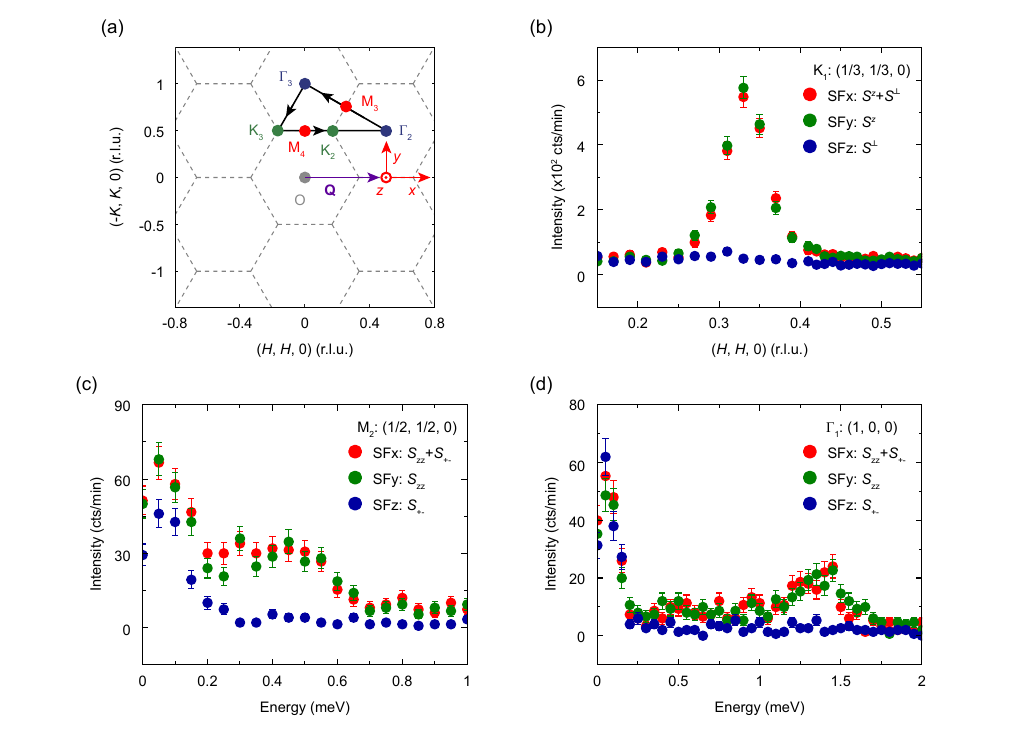}
\caption{Polarized neutron scattering measurements of a {\TMGO} single crystal at 0~T and 1.7~K. (a) Sketch of the reciprocal space in the ($H$, $K$, 0) plane. The red arrows indicate the polarization directions at a representative {\Q}. (b) Qscans near the magnetic Bragg peak K$_1$ point at $E$ = 0~meV. (c),(d) Energy scans at the high symmetry points, M$_2$ and $\Gamma_1$.}
\label{fig:pol}
\end{figure*}

The inelastic neutron scattering measurements were performed on the cold neutron multi-chopper spectrometer LET at the Rutherford Appleton Laboratory, Didcot, UK, and the AMATERAS cold neutron disk chopper spectrometer at the Japan Proton Accelerator Research Complex \cite{AMATERAS}. For the LET experiment, the incident energies are chosen to be 6, 3.15 and 1.94~meV with energy resolutions of 0.185, 0.069 and 0.031~meV, respectively. A dilution insert was used to reach the base temperature of 0.12~K, which was equipped with a vertical magnet to provide external fields. The same pieces of single crystals with a total mass of 17.2~g were used as in Ref.~\onlinecite{TMGO_Yao}, which were co-aligned in the ($H$, $K$, 0) plane. For the AMATERAS experiment, we choose the incident energy to be 2.6 and 5.9~meV and the corresponding energy resolutions are 0.045 and 0.155~meV,respectively. A dilution insert with a vertical magnet is used in this experiment. One piece of single crystal was aligned in the ($H$, $K$, 0) plane which could rotate around the vertical $z$ axis. The data were analysed using the Horace-Matlab suite \cite{HORACE}.

The polarized neutron scattering experiments were carried out on the ThALES cold triple-axis spectrometer at the Institut Laue-Langevin, Grenoble, France. In this experiment, we used a focusing Heusler monochromator and analyzer to produce and analyse the polarized neutrons with the fixed final energy of $E_\mathrm{f}$ = 4.662~meV. Longitudinal polarization analysis was done using the Helmholtz coils. A base temperature of 1.7 K is realized with this setup. One piece of single crystal with mass of $\sim$0.3~g was used for the measurement.

\section{Polarized neutron scattering measurements.}

For the polarized neutron scattering measurements, we used the conventional notations to define the polarization directions, $x$, $y$ and $z$, in reciprocal space. To be more specific, $x$ is defined in such a way that it keeps parallel to the momentum transfer {\Q}, and $y$ is perpendicular to $x$ but remains in the scattering plane while $z$ is perpendicular to both $x$ and $y$, thus perpendicular to the scattering plane (Fig.~\ref{fig:pol}a). During the measurements, we polarized the incident neutrons along one of the basic directions, $x$, $y$ or $z$, and selectively measured the scattered neutrons parallel or anti-parallel to the incident neutrons, which are marked as {\gls*{NSF}} and {\gls*{SF}} channels, respectively. Correspondingly, there are six different channels in total, {\SFx}, {\SFy}, {\SFz}, {\NSFx}, {\NSFy} and {\NSFz}.

\begin{figure*}[bt]
\includegraphics[width=\linewidth]{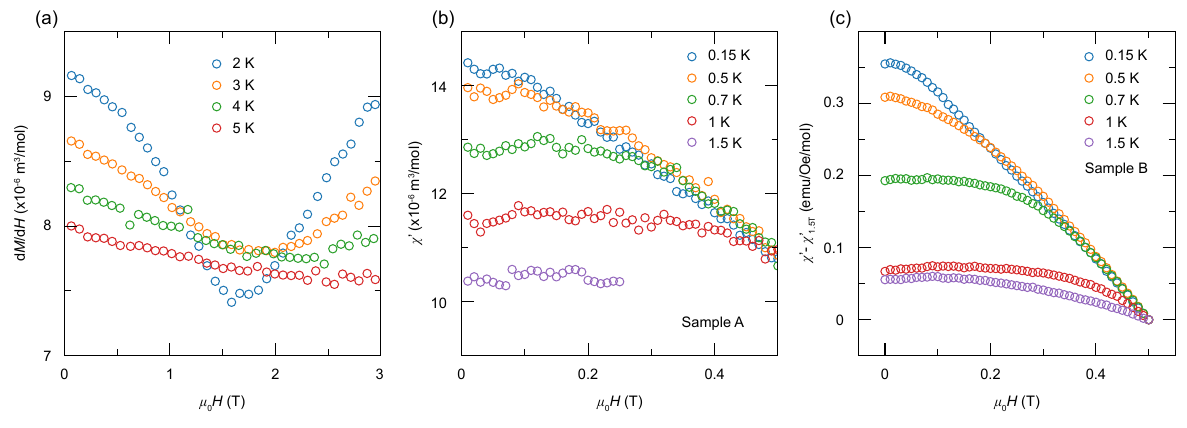}
\caption{Magnetic susceptibility in magnetic fields along the $z$ axis. (a) Field dependence of the DC differential susceptibility d$M$/d$H$ above 2~K. (b), (c) Real part of the ac susceptibility $\chi$' as a function of external field at different temperatures. The measurement was performed with ac frequency of 469 Hz on two different samples. No divergence or enhancement related to the {\gls{BKT}} phase is observed.}
\label{fig:sus}
\end{figure*}

\begin{figure*}[bt]
\includegraphics{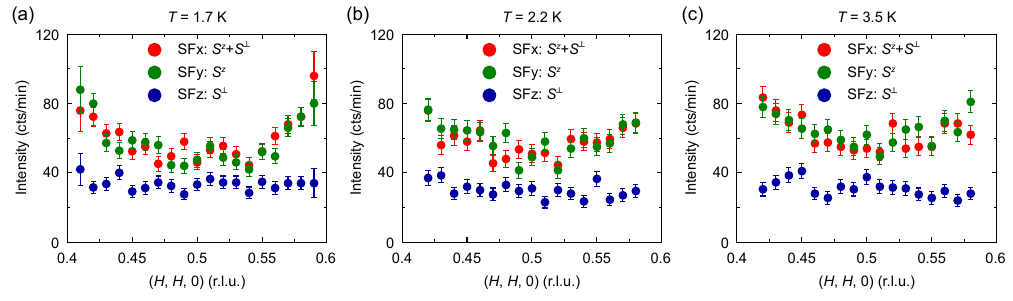}
\caption{Polarized neutron scattering measurements around the M$_2$ point at the indicated temperatures and $E$ = 0~meV.}
\label{fig:M}
\end{figure*}

\begin{figure*}[bth]
\includegraphics{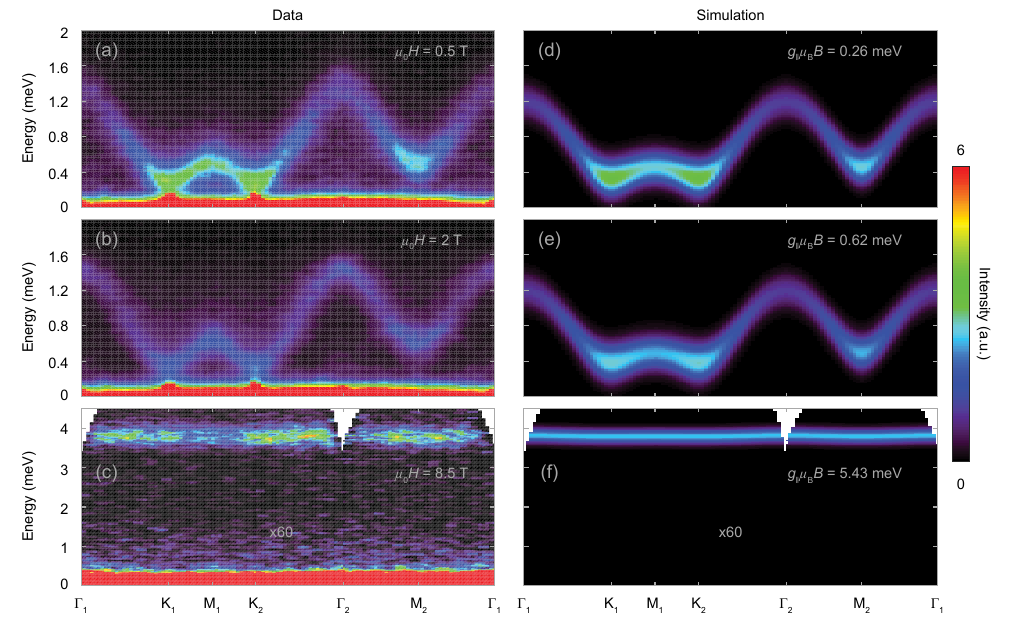}
\caption{Spin wave dispersion in different external longitudinal fields measured at LET and 0.12~K. (a)-(c) Spin excitations along the high symmetry reciprocal directions marked by the black arrows in Fig.~1b in the indicated fields. (c)-(f) Calculated spin wave dispersions. The intensities in (c) and (f) are multiplied by 60.}
\label{fig:LET}
\end{figure*}

\begin{figure*}[bth]
\includegraphics{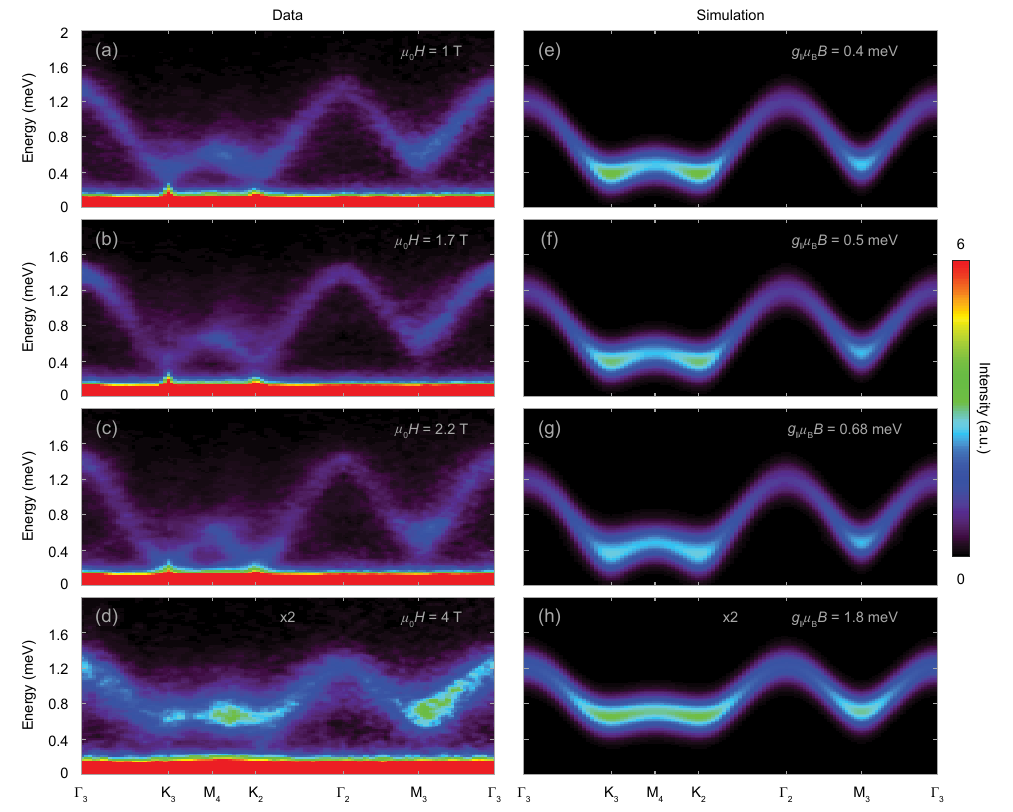}
\caption{Spin wave dispersion in different external fields at 0.06~K. (a)-(d) Spin excitations along the high symmetry reciprocal directions marked by the black arrows in Fig.~S1a in the indicated fields. (e)-(h) Calculated spin wave dispersions. The intensities in (d) and (h) are multiplied by 2.}
\label{fig:AMATERAS}
\end{figure*}

\begin{figure*}[bt]
\includegraphics{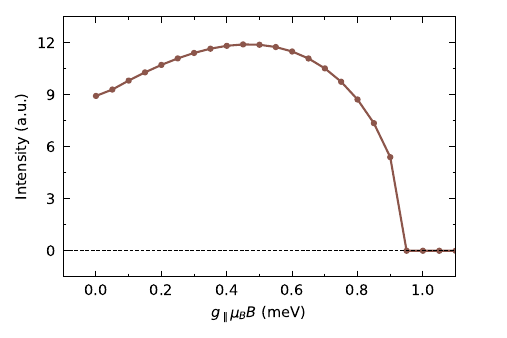}
\caption{Calculated field dependent evolution of magnetic Bragg peak intensity at {\Q} = (1/3, 1/3, 0) using {\gls*{LSW}}.}
\label{fig:calc}
\end{figure*}

In the neutron scattering measurements, the neutrons are only sensitive to the magnetic components perpendicular to {\Q}. Moreover, the spin of a neutron can only be flipped by magnetic components perpendicular to it with a flipping ratio determined by the instrument, leading to leakage of the signal from {\gls*{SF}} channel to {\gls*{NSF}} channel and vice versa. For elastic measurements, the scattering process probes the static spins along the $z$ direction ($S^z$) or perpendicular to $z$ ($S^\perp$) while the inelastic measurements detect the spin-spin correlations out of the basic plane ($S_{zz}$) and within the plane ($S_{+-}$). In the configuration of our experiments, the {\SFx} channel corresponds to the components that are perpendicular to {\Q}, which are $S^z$/$S_{zz}$ and part of $S^\perp$/$S_{+-}$. Meanwhile, the {\SFy} channel contains the components that are perpendicular to both {\Q} and $y$, which are $S^z$/$S_{zz}$ only, and the {\SFz} channel detects the components in the $xy$ plane but perpendicular to {\Q}, which are part of $S^\perp$/$S_{+-}$ but no $S^z$/$S_{zz}$.

We start by the polarization analysis of the elastic signals. In Fig.~\ref{fig:pol}b, we present the momentum scans across the magnetic Bragg peak K$_1$ in all three {\gls*{SF}} channels. The measurements were performed at 1.7~K, which is above the magnetic peak saturating temperature ($\sim$0.4~K). At this temperature range, the linewidth of the excitation spectrum is slightly broadened, but the overall spin excitation dispersion should be still similar to that observed at 0.05~K \cite{TMGO_Yao}. With signals absent in {\SFz} channel, the peaks in {\SFx} and {\SFy} channels are comparable which means that only $S^z$ component is observed in elastic neutron measurements, consistent with our model that the transverse $S^\perp$ component behaves as multipoles and cannot be detected directly. Moreover, we performed the constant {\Q} scans at the high symmetry points, K$_1$, M$_2$ and $\Gamma_1$, and only magnetic excitations of $S_{zz}$ components are present in these scans (Fig.~\ref{fig:pol}c,~d, Fig.~4a). The peak-like features close to the zero energy in Fig.~\ref{fig:pol}c,~d may come from the background or leakage since they do not obey the fundamental relationship: {\SFy} + {\SFz} = {\SFx}. For conventional localized systems, the magnetic excitations run perpendicular to the static spins. Here, the spin dynamics occur in the longitudinal channels, indicating quantum effect and multipolar nature of the transverse spin components.

\section{Susceptibility measurements.}

It has been theoretically proposed that in the {\gls*{TFIM}} on a triangular lattice, the low-temperature three-sublattice ground state will melt in a two-step manner with an intermediate {\gls*{BKT}} phase \cite{TFIM_Isakov, TFIM_Damle, TFIM_Biswas, TMGO_WLi}. In {\TMGO}. It has been predicted that the uniform susceptibility will diverge in an external field along the $c$ direction in the {\gls*{BKT}} phase \cite{TFIM_Damle, TFIM_Biswas}. According to the theoretical calculations, the susceptibility tends to diverge as $\chi(B) \sim |B|^{-[(4-18\eta)/(4-9\eta)]}$ where $\eta(T) \in (1/9, 2/9)$ (Ref.~\onlinecite{TFIM_Damle, TFIM_Biswas}). To test this result, we measured the magnetic susceptibility as a function of magnetic fields and temperatures (Fig.~\ref{fig:sus}). No divergence in susceptibility was observed in the zero field limit. We note that the compressible {\gls*{BKT}} phase for the {\gls*{TFIM}}, that is characterized by a divergent correlation length and quasi-long-range order in a temperature window above the true long-range ordering transition and describes the long-distance and low-energy properties of the {\gls*{TFIM}}, is fragile to inter-layer coupling or weak disorder.

In addition, Li \textit{et al.} suggested that there would be a magnetic peak at M points at finite temperature due to the proliferated vortex-antivortex pairs based on Monte-Carlo calculations \cite{TMGO_WLi}. We performed polarized elastic neutron scattering measurement across M$_2$ points at different temperatures and no magnetic peaks were found (Fig.~\ref{fig:M}).

\section{More inelastic neutron scattering data.}

The inelastic neutron scattering measurements were performed in different external longitudinal fields, 0, 0.5, 1.5, 2, 2.5, 3, 5, 8.5~T for experiments at LET and 0, 1, 1.7, 2.2, 4~T for experiments at AMATERAS. For the measurements at LET, the 8.5~T data can be taken as an ideal background since the spin gap is high enough to disentangle the low-energy background and the high-energy signals. Thus, in Fig.~4b, for the data measured between 0 and 5~T we show the background subtracted results while the raw 8.5~T data is presented. By integrating the scattering function along the energy axis, we can determine the total local moment, $\langle m^2\rangle$, which are shown as \textit{All} in Fig.~4c. The \textit{elastic} contributions are evaluated by fitting the elastic lines with Gaussian profiles. The difference between the total moment and elastic channel gives the \textit{inelastic} spectra. For the experiments at AMATERAS, no high-field data are available for background subtraction.

In the main text, we have shown the dispersion under several representative external fields (Fig.~3), along with the spin wave calculation using the \textsc{spinw} program \cite{SPINW}. Here, we present the rest of the data, including the LET data under 0.5, 2 and 8.5~T (Fig.~\ref{fig:LET}) and AMATERAS data under 1, 1.7, 2.2 and 4~T (Fig.~\ref{fig:AMATERAS}). The overall dispersions are well reproduced by {\gls*{LSW}} simulations. Furthermore, we use the same approach and parameters to calculate the field dependent magnetic Bragg peak intensity at {\Q} = (1/3, 1/3, 0). Only $S^z$ is considered due to multipolar nature of $S^x$ and $S^y$. As indicated by Fig.~\ref{fig:calc}, the peak intensity shows a non-monotonic behavior with increasing field, consistent with our data. Due to the stronger quantum fluctuations around zero field which is not fully caught my our mean-field approach, the measured peak intensity is weaker at low field compared with calculation.